\documentclass[preprint]{elsarticle}

\usepackage{amsmath}
\usepackage{amssymb}
\usepackage{graphicx}
\usepackage{subfigure}
\usepackage{algorithm}
\usepackage{algorithmic}
\usepackage{natbib}

\usepackage{lineno,hyperref}

\journal{Signal Processing}

\begin{document}

\begin{frontmatter}

\title{MIMO radar waveform design with practical constraints: A low-complexity approach}



\author[mymainaddress]{Chenglin Ren\corref{mycorrespondingauthor}}
\cortext[mycorrespondingauthor]{Corresponding author}
\ead{rencl3@bit.edu.cn}

\author[mymainaddress]{Fan liu}
\author[mysecondaryaddress]{Longfei Zhou}
\author[mymainaddress]{Jianming Zhou}
\author[mysecondaryaddress]{Wu Luo}
\author[mymainaddress]{Shengzhi Yang}


\address[mymainaddress]{School of Information and Electronics, Beijing Institute of Technology, Beijing 100081, China}
\address[mysecondaryaddress]{State Key Laboratory of Advanced Optical
Communication Systems and Networks, Peking University, Beijing 100871,
China}

\begin{abstract}
In this letter, we consider the multiple-input multiple-output (MIMO) radar waveform design in the presence of signal-dependent clutters and additive white Gaussian noise. By imposing the constant modulus constraint (CMC) and waveform similarity constraint (SC), the signal-to-interference-plus-noise (SINR) maximization problem is non-convex and NP-hard in general, which can be transformed into a sequence of convex quadratically constrained quadratic programming (QCQP) subproblems. Aiming at solving each subproblem efficiently, we propose a low-complexity method termed Accelerated Gradient Projection (AGP). In contrast to the conventional IPM based method, our proposed algorithm achieves the same performance in terms of the receive SINR and the beampattern, while notably reduces computational complexity.
\end{abstract}


\end{frontmatter}


\section{Introduction}
MIMO radar has been intensively studied as a new paradigm of the radar system \cite{1597550,4408448,1399143,4350230}. By allowing individual waveforms to be transmitted at each antenna, MIMO radar is able to exploit extra degrees of freedom in contrast to its phased-array counterpart, and therefore achieves a more favorable performance. In the existing literature \cite{4516997,6194367,5605365,6649991}, radar waveform optimization can be classified into two categories: waveform design by only considering the radar transmitter \cite{4516997,6194367} and waveform design by jointly considering radar transmitter and receive filter \cite{5605365,6649991}, where we focus on the second category.

Pioneered by the exploration of \cite{4383615}, the transmit waveform is designed by maximizing the receive SINR in the presence of the signal-dependent clutters and the Gaussian noise. In practice, for the use of non-linear power amplifiers in the numerous radar systems, the CMC is typically involved, which is important to the SINR performance of the radar \cite{4383615}. In addition, the SC enables a flexible tradeoff between the output SINR and the desired autocorrelation properties by controlling the similarity between the reference waveform and its optimized counterpart. However, the waveform design with both CMC and SC leads to the non-convexity and the NP-hardness of the problem. To overcome such a challenge, the previous works \cite{7060447,7450660} either omit or relax these constraints in favour of the existing solvers, which result in suboptimal solutions and high computational costs.

Inspired by the recent research \cite{7450660}, where the non-convex waveform design problem is relaxed to a sequence of convex QCQP subproblems, we develop in this letter a different iterative relaxing scheme to produce the sequential QCQP subproblems and a novel AGP algorithm to solve each subproblem. Furthermore, the proposed AGP involves two parts: the fast iterative shrinkage-thresholding algorithm (FISTA) procedure and the customized projection procedure, where the former is based on the convex optimization theory and the latter is highly dependent on the specific feasible region formulated by the CMC and SC. In contrast to the conventional IPM that is employed in \cite{7450660}, our proposed algorithm obtains a comparable performance in terms of the receive SINR as well as the beampattern. In addition, the AGP shares a much lower computational complexity, and directly solves the complex convex QCQP subproblems without converting them into the real representations \cite{7450660}.

\section{System Model}
We consider a colocated narrow band MIMO radar system with ${N_T}$ transmit antennas and ${N_R}$ receive antennas. We suppose that ${\mathbf{t}} = {[{\mathbf{t}}_1^T, ... ,{\mathbf{t}}_N^T]^T} \in {\mathbb{C}}^{N_TN \times 1}$ is the vectorized transmit wavform, where ${( \cdot )^T}$ denotes the transpose, $N$ is the number of samples, and ${{\mathbf{t}}_n} \in {\mathbb{C}^{{N_T} \times 1}}$, $n = 1,...,N$, stands for the $n$-th sample across the ${N_T}$ antennas. Then the receive waveform ${\mathbf{r}} \in \mathbb{C}^{N_RN \times 1}$ is given by \cite{6649991}
\begin{equation}\label{1}
{\mathbf{r}} = {\alpha _0}{\mathbf{M}}({\phi _0}){\mathbf{t}} + \sum\limits_{m = 1}^M {{\alpha _m}{\mathbf{M}}({\phi _m}){\mathbf{t}} + {\mathbf{n}}}
\end{equation}
where ${\mathbf{n}} \in {\mathbb{C}}^{{N_R}N \times 1}$ stands for the circular complex white Gaussian noise with zero mean and covariance matrix $\sigma _n^2{\mathbf{I}}$, ${\alpha _0}$ and ${\alpha _m}$ represent the complex amplitudes of the target and the $m$-th interference source, ${\phi _0}$ and ${\phi _m}$ are the angle of the target and the angle of the $m$-th interference source, respectively, ${\mathbf{M}}(\phi )$ denotes the steering matrix of a Uniform Linear Array (ULA) with half-wavelength separation between the antennas, which is given by
\begin{equation}\label{2}
{\mathbf{M}}(\phi ) = {{\mathbf{I}}_N} \otimes [{{\mathbf{a}}_r}(\phi ){{\mathbf{a}}_t}{(\phi )^T}]
\end{equation}
where ${{\mathbf{I}}_N}$ is the $N \times N$ identity matrix, $\otimes$ denotes the Kronecker product, ${{\mathbf{a}}_t}$ and ${{\mathbf{a}}_r}$ are the transmit and the receive steering vectors, respectively.

As mentioned above, aiming for maximizing the output SINR, we propose to jointly optimize the transmit waveform and the receive filter. Without loss of generality, we employ a linear Finite Impulse Response (FIR) filter $ \mathbf{f} \in \mathbb{C}^{N_RN \times 1}$ to process the received echo wave. This is given as
\begin{equation}\label{3}
r = {{\mathbf{f}}^H}{\mathbf{r}} = {\alpha _0}{{\mathbf{f}}^H}{\mathbf{M}}({\phi _0}){\mathbf{t}} + \sum\limits_{m = 1}^M {{\alpha _m}{{\mathbf{f}}^H}{\mathbf{M}}({\phi _m}){\mathbf{t}}}  + {{\mathbf{f}}^H}{\mathbf{n}}
\end{equation}
where ${( \cdot )^H}$ denotes the Hermitian transpose. As a result, the output SINR can be expressed as
\begin{equation}\label{4}
\text{SINR} = \frac{{\sigma {{\left| {{{\mathbf{f}}^H}{\mathbf{M}}({\phi _0}){\mathbf{t}}} \right|}^2}}}{{{{\mathbf{f}}^H}{\mathbf{\tilde S}}(t){\mathbf{f}} + {{\mathbf{f}}^H}{\mathbf{f}}} }
\end{equation}
where $\sigma  = E[{\left| {{\alpha _0}} \right|^2}]/\sigma _n^2$, with $E[ \cdot ]$ being the statistical expectation, and
\begin{equation}\label{5}
{\mathbf{\tilde S}}(t)  = \sum\limits_{m = 1}^M {{I_m}{\mathbf{M}}({\phi _m}){\mathbf{t}}{{\mathbf{t}}^H}{{\mathbf{M}}^H}({\phi _m})}
\end{equation}
where ${I_m} = E[{\left| {{\alpha _m}} \right|^2}]/\sigma _n^2$.

\section{Problem Formulation}
Following the approach in \cite{7450660}, we incorporate the CMC and SC in the maximization of the output SINR, resulting in an optimization problem as follows
\begin{equation}\label{6}
\begin{gathered}
  \mathop {\max }\limits_{{\mathbf{f}},{\mathbf{t}}}\;\;\; \frac{{\sigma {{\left| {{{\mathbf{f}}^H}{\mathbf{M}}({\phi _0}){\mathbf{t}}} \right|}^2}}}{{{{\mathbf{f}}^H}{\mathbf{\tilde S}}(t){\mathbf{f}} + {{\mathbf{f}}^H}{\mathbf{f}}}} \hfill \\
  s.t.\;\;\;\;\;\left| {t(k)} \right| = 1/\sqrt {{N_T}N} \hfill \\
  \;\;\;\;\;\;\;\;\;\;\;{\left\| {{\mathbf{t}} - {{\mathbf{t}}_0}} \right\|_\infty } \le \varepsilon  \hfill \\
\end{gathered}
\end{equation}
where ${{t}}(k)$ is the $k$-th entry of ${\mathbf{t}}$, $k = 1,...,{N_T}N$, ${\left\|  \cdot  \right\|_\infty }$ denotes the infinity norm, and ${{\mathbf{t}}_0}$ represents the reference waveform. By taking into account the CMC, the SC can be rewritten as
\begin{equation}\label{7}
\arg {{t}}(k) \in [{\omega _k},{\omega _k} + \delta ]
\end{equation}
where ${\omega _k}$ and ${\delta}$ are respectively given by
\begin{equation}\label{8}
\begin{gathered}
  {\omega _k} = \arg {{{t}}_0}(k) - arccos(1 - {\varepsilon ^2}/2) \hfill \\
  \delta = 2\arccos (1 - {\varepsilon ^2}/2) \hfill \\
\end{gathered}
\end{equation}
\noindent where ${{t_0}}(k)$ is the $k$-th entry of ${\mathbf{t}_0}$, and the prespecified parameter $\varepsilon$ ($0 \le \varepsilon  \le 2$) stands for the similarity between ${\mathbf{t}}$ and ${{\mathbf{t}}_0}$. In addition, noting that ${\mathbf{f}}$ is unconstrained \cite{6649991}, the optimization of ($6$) is equivalent to
\begin{equation}\label{9}
\begin{gathered}
  \mathop {\max }\limits_{\mathbf{t}}\;\;\; {{\mathbf{t}}^H}{\mathbf{\Psi }}({\mathbf{t}}){\mathbf{t}} \hfill \\
  s.t.\;\;\;\;\;\left| {t(k)} \right| = 1/\sqrt {{N_T}N}  \hfill \\
  \;\;\;\;\;\;\;\;\;\;\;\arg t(k) \in [{\omega _k},{\omega _k} + \delta ] \hfill \\
\end{gathered}
\end{equation}
where ${\mathbf{\Psi}} ({\mathbf{t}})$ is a positive-semidefinite matrix in terms of the output SINR and is given by
\begin{equation}\label{10}
{\mathbf{\Psi}} ({\mathbf{t}}) = {{\mathbf{M}}^H}({\phi _0}){[{\mathbf{\tilde S}}(t)  + {\mathbf{I}}]^{ - 1}}{\mathbf{M}}({\phi _0}).
\end{equation}
According to the previous study \cite{4383615}, it is possible to obtain a suboptimal SINR by assuming $ {\mathbf{\Psi}}= {\mathbf{\Psi}} ({\mathbf{t}})$ with a fixed ${\mathbf{t}}$ and optimizing ${\mathbf{t}}$ with the new ${\mathbf{\Psi}}$ iteratively. However, even for a fixed ${\mathbf{\Psi}}$, the optimization of ${\mathbf{t}}$ is still non-convex and NP-hard due to the equality constraints involved. In the next section, by referring to the method of Successive QCQP Refinement---Binary Search (SQR-BS) \cite{7450660}, we introduce a low-complexity algorithm to solve ($9$).

\section{Proposed Algorithm}
As analyzed above, we denote ${\mathbf{P}} = ({\mathbf{\Psi}}  - \lambda\mathbf{I})$, where $\mathbf{I}$ is a identity matrix, $\lambda  \ge {\lambda _{\max }}({\mathbf{\Psi}})$ and ${\lambda _{\max }}({\mathbf{\Psi}} )$ is the maximum eigenvalue of ${\mathbf{\Psi}}$, which guarantees that ${\mathbf{P}}$ is negative-semidefinite. Hence, the optimization problem of ($9$) is equivalent to the following non-convex problem
\begin{equation}\label{11}
\begin{gathered}
  \mathop {\max }\limits_{\mathbf{t}}\;\;\; {{\mathbf{t}}^H}{\mathbf{Pt}} \hfill \\
  s.t.\;\;\;\;\;\left| {{{t}}(k)} \right| = 1/\sqrt {{N_T}N} \hfill \\
  \;\;\;\;\;\;\;\;\;\;\;\arg {{t}}(k) \in [{\omega _k},{\omega _k} + \delta ]. \hfill \\
\end{gathered}
\end{equation}
It is easy to see that the feasible region for each entry of $\mathbf{t}$ can be viewed as a circular arc from A to B as shown in Figure 1.
\begin{figure*}[ht]
\centering
\includegraphics[width=0.5\columnwidth]{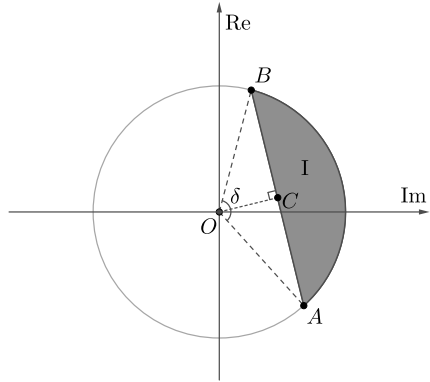}

\caption{Feasible region for each dimension.}
\label{1}
\end{figure*}
For notational convenience, we denote ${x_k} = \operatorname{Re}({{t}}(k))$, ${y_k} = \operatorname{Im} ({{t}}(k))$. The optimization problem of ($11$) can be relaxed as the following convex QCQP problem
\begin{equation}\label{12}
\begin{gathered}
  \mathop {\max }\limits_{\mathbf{t}}\;\;\; {{\mathbf{t}}^H}{\mathbf{Pt}} \hfill \\
  s.t.\;\;\;\;\;\left| {{{t}}(k)} \right| \le 1/\sqrt {{N_T}N} \hfill \\
  \;\;\;\;\;\;\;\;\;\;\;{c_x}{x_k} + {c_y}{y_k} \ge c_x^2 + c_y^2 \hfill \\
\end{gathered}
\end{equation}
where
\begin{equation}\label{13}
\begin{gathered}
  {c_x} = \frac{{\cos ({\omega _k}) + cos({\omega _k} + \delta )}}{2} \hfill \\
  {c_y} = \frac{{sin({\omega _k}) + \sin ({\omega _k} + \delta )}}{2} \hfill \\
\end{gathered}
\end{equation}
represent the abscissa and the ordinate of the point C, respectively, as shown in Figure 1. The feasible region of each ${{t}}(k)$ is labeled by zone ${\rm I}$, which is the convex hull composed by the circular arc and the line segment.

Noting that the relaxation becomes closer to ($11$) as the value of $\delta $ becomes smaller, the author of \cite{7450660} reduces $\delta $ by half and equally divides the feasible region into two parts. By iteratively selecting the part in which the optimized $\mathbf{t}$ locates and fixing ${\mathbf{P}}$ with a given ${\mathbf{t}}$ as analyzed in the last section, the sequential convex QCQP subproblems in the form of ($12$) are obtained. The number of times of dividing is given in advance, and each dividing is called one refinement where the result is iteratively calculated via the conventional IPM, which we refer readers to \cite{7450660} for more details.

For clarity, here we note the following two remarks:
\begin{enumerate}[(1)]
\item \emph{Remark 1}: Compared with the SQR-BS algorithm in \cite{7450660}, we do not fix the number of the refinements which is equal to iterations, and reduce $\delta $ until it is sufficiently small.
\item \emph{Remark 2}: It is worth highlighting that the feasible region in ($12$) is convex, which enables a closed-form projector to be used in the gradient projection method. We hereby propose the AGP approach that involves of two steps: the FISTA procedure and the projection procedure. In the first step, we compute the gradient of the objective function, and obtain the descent direction by the FISTA method. Then we project the obtained point onto the feasible region by a specifically tailored orthogonal projector.
\end{enumerate}

In what follows, we discuss the details of the AGP algorithm.

\subsection{FISTA Procedure}

The FISTA procedure accelerates the iterative shrinkage-thresholding algorithm (ISTA) by introducing a smart interpolation factor, which leads to a faster convergence rate compared to that of the ISTA method \cite{doi:10.1137/080716542}. Both algorithms are computational-efficient for unconstrained convex optimization problems.

Firstly, we denote ${\mathbf{y}}({\mathbf{t}}) = {{\mathbf{t}}^H}{\mathbf{Pt}}$, and then the gradient of the objective function in ($12$) can be caculated as
\begin{equation}\label{14}
{\mathbf{g}}({\mathbf{t}}) = {\nabla _{\mathbf{t}}}{\mathbf{y}}({\mathbf{t}}) = ({\mathbf{P}} + {{\mathbf{P}}^H}){\mathbf{t}} = 2{\mathbf{Pt}}
\end{equation}
since ${\mathbf{P}}$ is a Hermitian matrix. Secondly, we derive the Hessian matrix as
\begin{equation}\label{15}
{\mathbf{H}} = \frac{{{\partial ^2}{\mathbf{y}}({\mathbf{t}})}}{{\partial {\mathbf{t}}\partial {{\mathbf{t}}^H}}} = {\mathbf{P}} + {{\mathbf{P}}^H} = 2{\mathbf{P}}.
\end{equation}
Thirdly, we set the stepsize $\tau  = 1/{\lambda _{\max }}({\mathbf{H}})$, where ${\lambda _{\max }}({\mathbf{H}})$ is the maximum eigenvalue of ${\mathbf{H}}$, namely the Lipschitz constant of the objective function. As a result, the iterative procedure can be simply described as
\begin{equation}\label{16}
\begin{gathered}
  {{\mathbf{v}}^{(n)}} = {{\mathbf{t}}^{(n)}} + {c_k}({{\mathbf{t}}^{(n)}} - {{\mathbf{t}}^{(n - 1)}}) \hfill \\
  {{\mathbf{t}}^{(n + 1)}} = {{\mathbf{v}}^{(n)}} - \tau  \cdot {2\mathbf{P}}{{\mathbf{v}}^{(n)}} \hfill \\
\end{gathered}
\end{equation}
where ${c_k} = (k - 1)/(k + 2)$ is the interpolation factor, and ${\mathbf{v}}$ is the transitive vector.

\subsection{Projection Procedure}
As discussed above, we assume that the optimization is unconstrained, i.e., the feasible region is the whole complex plane for each t(k). Inspired by \cite{arxiv/1711.05220}, we design a projection procedure that projects each ${t}(k)$ onto the zone ${\rm I}$ as shown in Figure 2. In what follows, we derive the projector under the two possible cases, i.e., the angle $\delta$ corresponding to the zone ${\rm I}$ is less or greater than $\pi$.

\begin{figure*}[ht]
\centering
\subfigure[]{
\includegraphics[width=0.47\columnwidth]{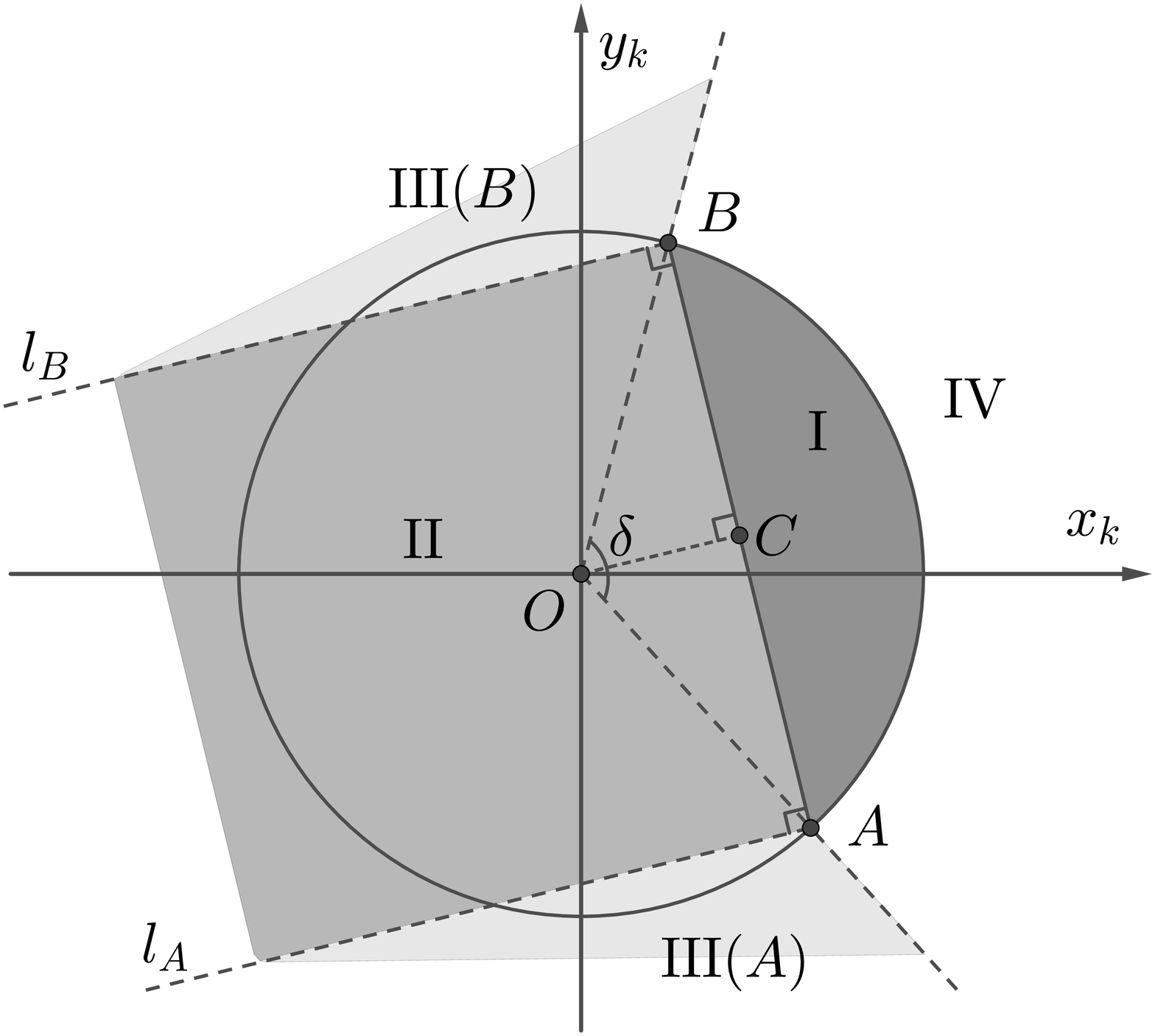}
}
\subfigure[]{
\includegraphics[width=0.47\columnwidth]{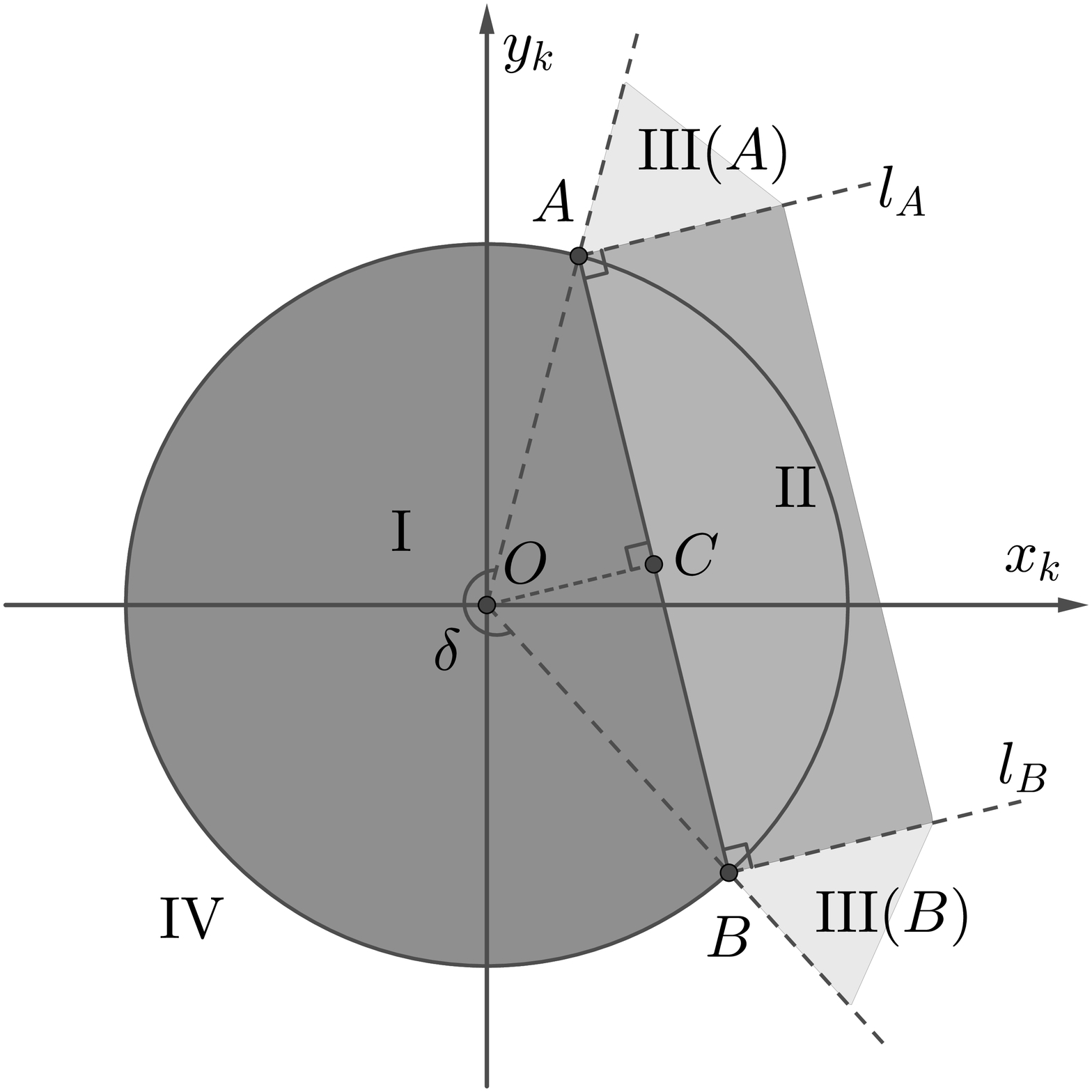}
}
\caption{Projection procedure for each dimension. (a) $\delta  \le \pi $; (b) $\delta  \ge \pi $. }
\label{2}
\end{figure*}

When $\delta  \le \pi $, as shown in Figure 2 (a), the real and the imaginary axes are labeled by ${x_k}$ and ${y_k}$, respectively. The two extreme points of zone ${\rm I}$ are marked with $A({a_x},{a_y})$ and $B({b_x},{b_y})$, respectively, given by
\begin{equation}\label{17}
\begin{gathered}
  {a_x} = \cos ({\omega _k}),{b_y} = \sin ({\omega _k}) \hfill \\
  {b_x} = \cos ({\omega _k} + \delta ),{a_y} = \sin ({\omega _k} + \delta ) \hfill \\
\end{gathered}
\end{equation}
\noindent and $C({c_x},{c_y})$ denotes the midpoint of line segment $AB$ as defined in ($13$). Then we draw two radial lines ${l_A}$ and ${l_B}$ which are both vertical to $AB$ and start at $A$ and $B$, respectively, and furthermore, by adding the radial lines ${OA}$ and ${OB}$, the whole complex plane ${\mathbb{C}}$ is divided into $5$ zones, i.e., ${\rm I}$, ${\rm I}{\rm I}$, ${\rm I}{\rm I}{\rm I}(A)$, ${\rm I}{\rm I}{\rm I}(B)$, ${\rm I}{\rm V}$. Given any $Q({x_k},{y_k}) \in {\mathbb{C}}$, we can obtain the nearest point $R(x_k^ * ,y_k^ * ) \in {\rm I}$ via the projection procedure. If $Q \in {\rm I}$, the projection is itself; if $Q \in {\rm I}{\rm I}$, $R$ is the foot of the perpendicular with $AB$ through $Q$; if $Q \in {\rm I}{\rm I}{\rm I}(A)$ or $Q \in {\rm I}{\rm I}{\rm I}(B)$, the nearest point is $A$ or $B$, respectively; if $Q \in {\rm I}{\rm V}$, $R$ is the normalization of $Q$. To sum up, the projection can be formulated as
\begin{equation}\label{18}
R = \left\{ \begin{gathered}
  Q,\;{c_x}{x_k} + {c_y}{y_k} \ge c_x^2 + c_y^2,\left| OQ \right| \le 1, \hfill \\
  F,\;{c_x}{x_k} + {c_y}{y_k} \le c_x^2 + c_y^2, \hfill \\
  \;\;\;\;\;{k_l}{x_k} + {d_B} \le {y_k} \le {k_l}{x_k} + {d_A}, \hfill \\
  A,\;{a_y}{x_k} - {a_x}{y_k} \le 0,{y_k} \ge {k_l}{x_k} + {d_A}, \hfill \\
  B,\;{b_y}{x_k} - {b_x}{y_k} \ge 0,{y_k} \le {k_l}{x_k} + {d_B}, \hfill \\
  Q/\left| OQ \right|,\;others \hfill \\
\end{gathered}  \right.
\end{equation}
where ${k_l} = ({a_y} + {b_y})/({a_x} + {b_x})$ represents the slope of ${l_A}$ and ${l_B}$, ${d_A} = {a_y} - {a_x}{k_l}$ and ${d_B} = {b_y} - {b_x}{k_l}$ represent the intercepts of ${l_A}$ and ${l_B}$, respectively, and  $F({x_f},{y_f})$ is the foot of the perpendicular on $AB$, which can be expressed as
\begin{equation}\label{19}
\begin{gathered}
  {x_f} = \frac{{c_x^2 + c_y^2 - {c_y}({y_k} - {k_l}{x_k})}}{{{c_x} + {k_l}{c_y}}} \hfill \\
  {y_f} = \frac{{c_x^2 + c_y^2 - {c_x}{x_f}}}{{{c_y}}}. \hfill \\
\end{gathered}
\end{equation}
\noindent Particularly, if ${c_x} = 0$, $F = ({x_k},{c_y})$; if ${c_y} = 0$, $F = ({c_x},{y_k})$.

When $\delta  \ge \pi $, as shown in Figure 2 (b), following the similar procedure of ($18$), the projection can be derived as
\begin{equation}\label{20}
R = \left\{ \begin{gathered}
  Q,\;{c_x}{x_k} + {c_y}{y_k} \le c_x^2 + c_y^2,\left| OQ \right| \le 1, \hfill \\
  F,\;{c_x}{x_k} + {c_y}{y_k} \ge c_x^2 + c_y^2, \hfill \\
  \;\;\;\;\;{k_l}{x_k} + {d_B} \le {y_k} \le {k_l}{x_k} + {d_A}, \hfill \\
  A,\;{a_y}{x_k} - {a_x}{y_k} \le 0,{y_k} \ge {k_l}{x_k} + {d_A}, \hfill \\
  B,\;{b_y}{x_k} - {b_x}{y_k} \ge 0,{y_k} \le {k_l}{x_k} + {d_B}, \hfill \\
  Q/\left| OQ \right|,\;others. \hfill \\
\end{gathered}  \right.
\end{equation}
where the only differences comparing with ($18$) are the signs of inequalities when $Q \in {\rm I}$ where $Q \in {\rm I}{\rm I}$, and $F({x_f},{y_f})$ can be also calculated by ($19$). For clarity, we summarize the proposed approach for solving the QCQP subproblems in Algorithm $1$.

\begin{algorithm}
\caption{Accelerated Gradient Projection (AGP)}
\label{alg:A}
\begin{algorithmic}
    \REQUIRE $\mathbf{P}$, ${\mathbf{t}}_0$, $\zeta$ (the desired threshold value), $\text{Num}$ (the maximum number of iterations), $\delta$ and ${\omega _k}, k = 1,...,{N_T}N$.
    \STATE  Compute ${\mathbf{g}}({\mathbf{t}})$, $t$,
    \STATE  Set $n = 1$, ${{\mathbf{t}}^{(1)}} = {{\mathbf{t}}^{(0)}} = {{\mathbf{t}}_0}$,
    \STATE  Compute ${{\mathbf{t}}^{(2)}}$ and it's projection ${{\mathbf{t}}^{(2)*}}$, Set $n = n + 1$,
    \WHILE{${{\mathbf{t}}^{(n ) * }} - {{\mathbf{t}}^{(n-1) * }} > \zeta $ and $n < \text{Num}$}
    \STATE  Compute ${{\mathbf{t}}^{(n + 1)}}$ by ($16$),
        \FOR{$k = 1,...,{N_T}N$}
        \STATE  Project ${{{t}}^{(n + 1)}}(k)$ to ${{{t}}^{(n + 1) * }}(k)$ with the given $\delta$, ${\omega _k}, k = 1,...,{N_T}N$ to get the projection ${{\mathbf{t}}^{(n + 1) * }}$
        \ENDFOR
        \STATE  Set $n = n + 1$.
    \ENDWHILE

    \ENSURE  ${{\mathbf{t}}^*} = 1/\sqrt {{N_T}N} \exp (j$arg$({{\mathbf{t}}^{(n)*}}))$ for problem ($11$).
\end{algorithmic}
\end{algorithm}

\emph{Remark 3}: The complexity of the AGP mainly comes from the computation of the FISTA in ($16$) with $3N_T^2{N^2} + 3{N_T}N$ complex flops and the tailored projection with $42{N_T}N$ complex flops, leading to a total computational complexity of ${\mathcal{O}}({{N}}_T^2{{{N}}^2})$ for each iteration. For converging to a suboptimal solution, the iteration number needed for the proposed algorithm is ${\mathcal{O}}({\log}(1/\zeta ))$ where $\zeta$ is the desired threshold value. On the other hand, the computational costs for the IPM is ${\mathcal{O}}({N}_T^{3.5}{{N}^{3.5}})$ \cite{6407265} and the number of iterations is ${\mathcal{O}}({\log}(1/\zeta ))$ as well \cite{5447068}.

\section{Numerical Results}
In this section, we provide numerical results to evaluate the performance in terms of the SINR and the beampattern $P(\phi )$ between the AGP and the IPM, and use chirp waveform as our benchmark. The reference waveform ${{\mathbf{t}}_0} \in \mathbb{C}^{{N_T}N \times 1}$ can be obtained by stacking the columns of ${{\mathbf{T}}_0}$, which is the orthogonal chirp waveform matrix defined as
\begin{equation}\label{24}
{{{T}}_0}(k,n) = \frac{{\exp [j2\pi k(n - 1)/N]\exp[j\pi {{(n - 1)}^2}/N]}}{{\sqrt {{N_T}N} }}
\end{equation}
where $k = 1,...,{N_T}$, $n = 1,...,N$. We assume the number of samples ${{N}} = 16$, and the number of transmit and the receive antennas ${{{N}}_T} = 4$, ${{{N}}_R} = 8$ respectively. In addition, we consider a scenario with three fixed signal-dependent clutters and additive white Gaussian disturbance with variance ${\sigma _n} = 0$dB. The power of the target echo and the three interfering sources are ${\left| {{\alpha _0}} \right|^2} = 10$dB, ${\left| {{\alpha _1}} \right|^2} = {\left| {{\alpha _2}} \right|^2} = {\left| {{\alpha _3}} \right|^2} = 30$dB, respectively, and the angle of the target and the three interferences are ${\phi _0} = {15^ \circ }$, ${\phi _1} = {-50^ \circ }$, ${\phi _2} = {-10^ \circ }$, ${\phi _3} = {40^ \circ }$, respectively.

\begin{figure*}[ht]
\centering
\subfigure[]{
\includegraphics[width=0.47\columnwidth]{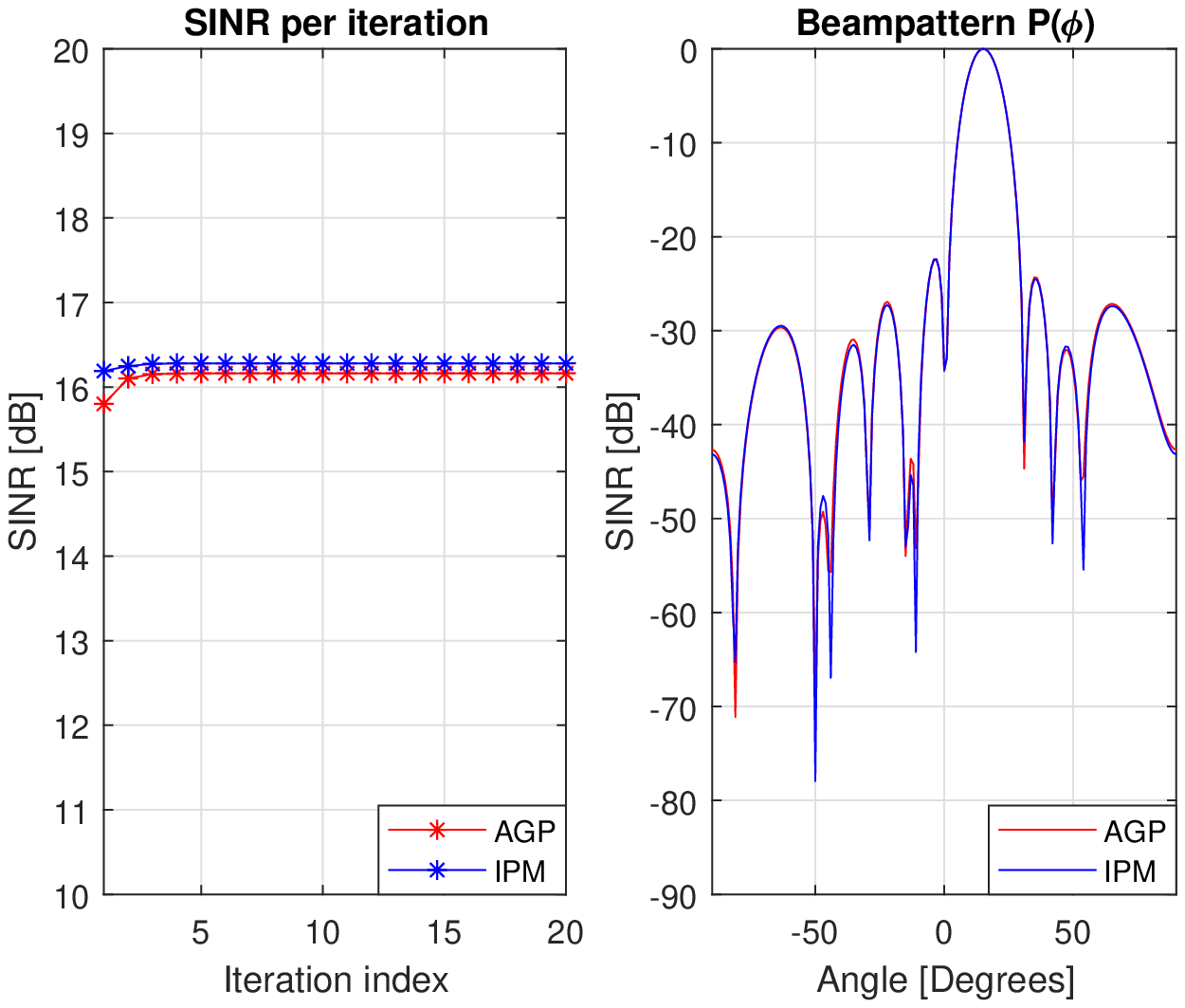}
}
\hspace{0cm}
\subfigure[]{
\includegraphics[width=0.47\columnwidth]{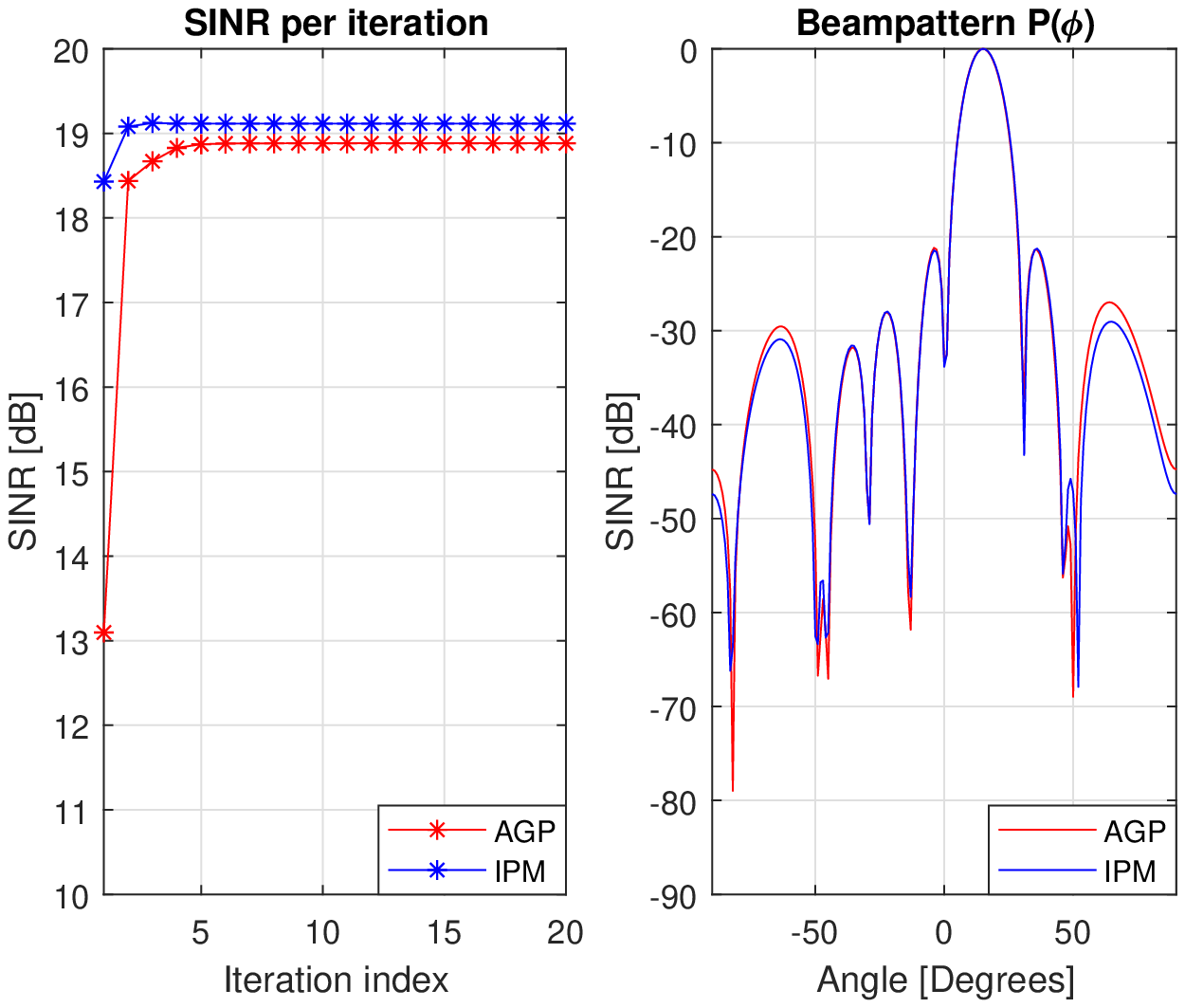}
}

\caption{The SINR in each iteration and the beampattern $P(\phi )$ of the optimal waveform for the AGP and the IPM approaches: (a) $\varepsilon  = 0.4$; (b) $\varepsilon  = 1.2$. }
\label{3}
\end{figure*}

\begin{figure*}[ht]
\centering
\subfigure[]{
\includegraphics[width=0.47\columnwidth]{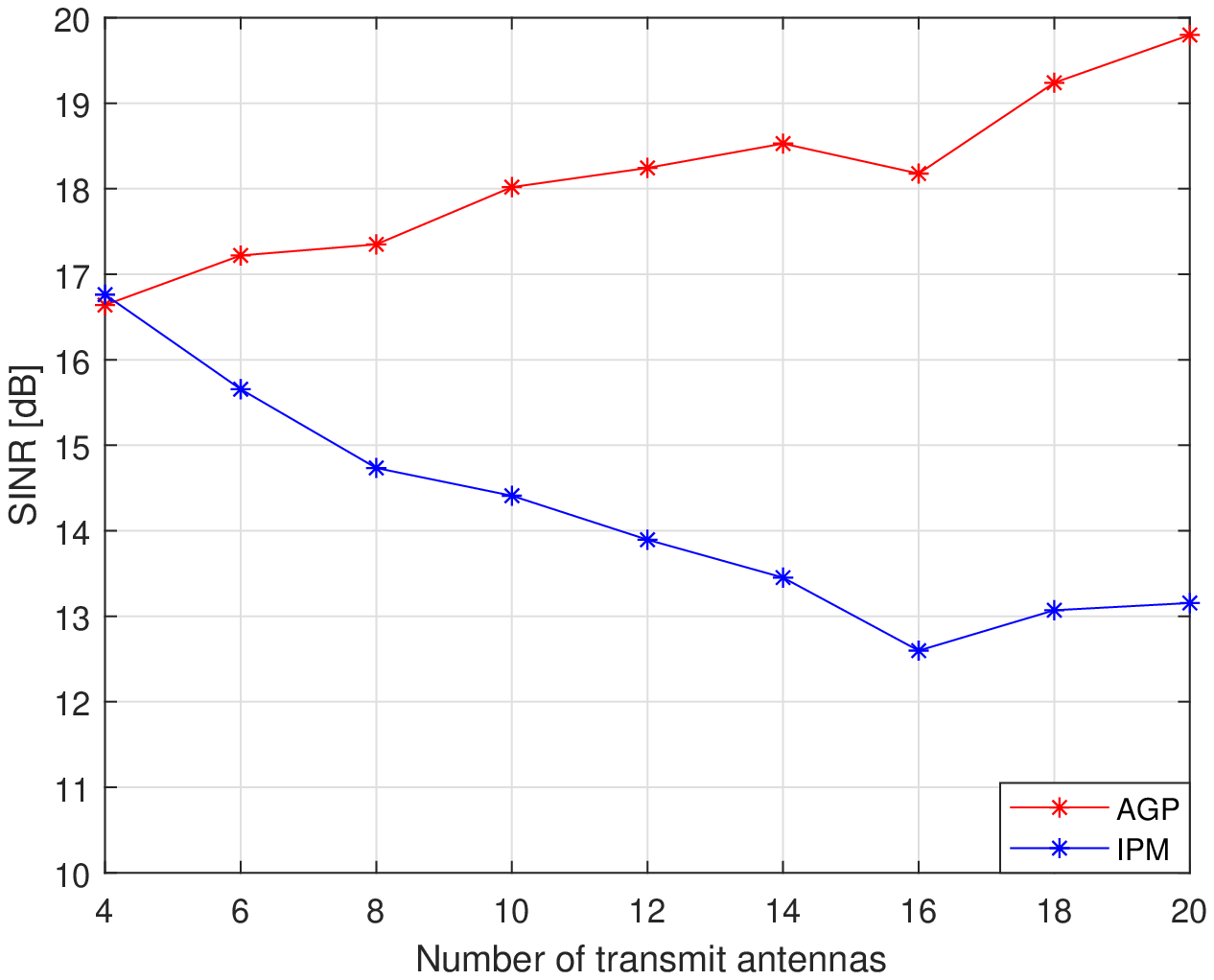}
}
\hspace{0cm}
\subfigure[]{
\includegraphics[width=0.47\columnwidth]{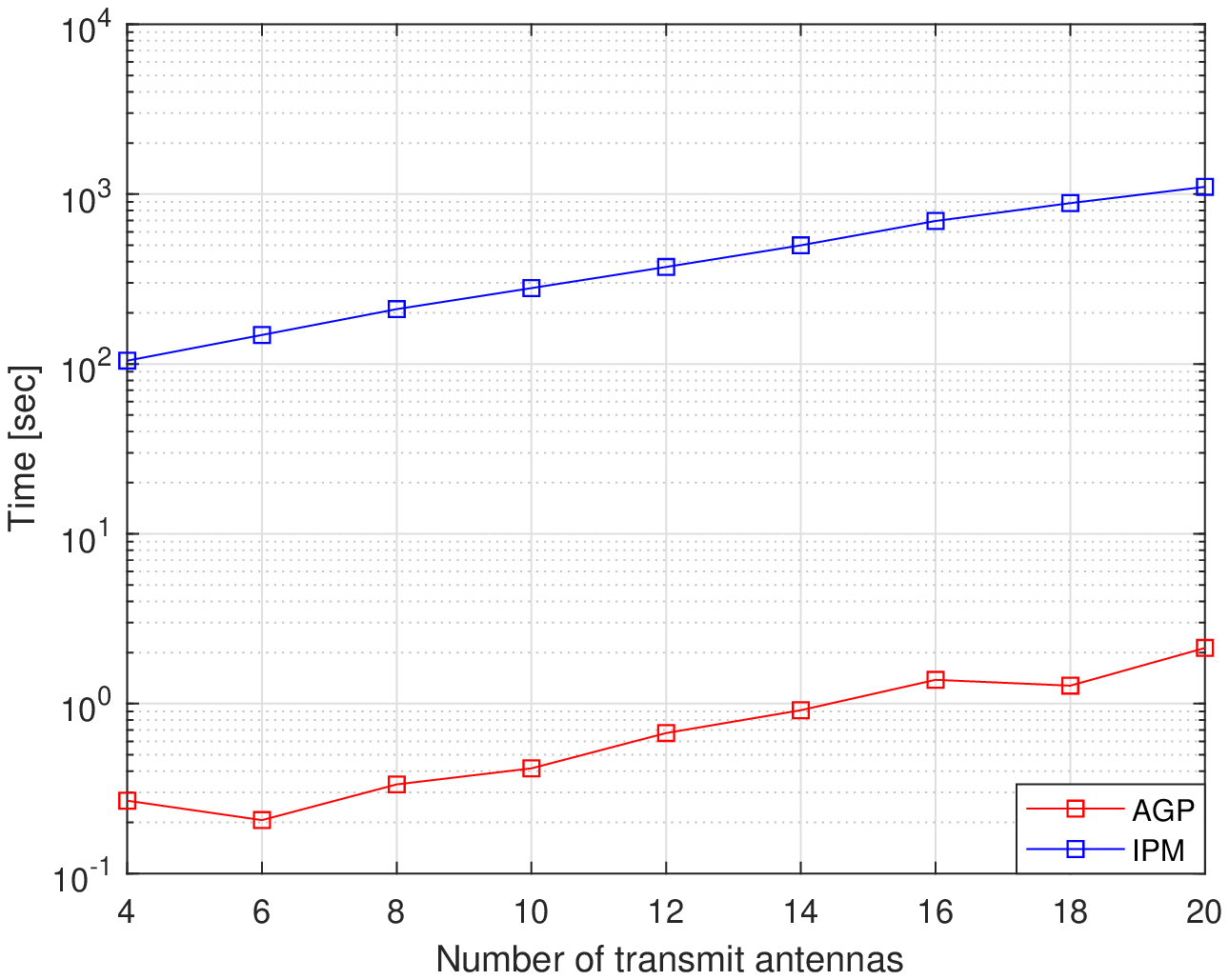}
}

\caption{(a) The SINR with increasing transmit antennas. (b) The CPU time with increasing transmit antennas. }
\label{4}
\end{figure*}

We first show the performance obtained by the two approaches in Figure 3 in terms of the SINR and the beampattern. When $\varepsilon  = 0.4$, Figure 3 (a) shows very close SINR results of the both approaches, while the beampattern resulting form the IPM exhibits better suppression performance. As the similarity loose, the beampattern of the AGP outperform the IPM when $\varepsilon  = 1.2$ as shown in Figure 3 (b), and the difference of the SINR is less than 0.25dB.

We further compare the SINR and the execution time of increasing transmit antennas for both approaches in Figure 4, where all the parameters remain the same with that of Figure 3 but $\varepsilon  = 0.5$. Figure 4 (a) indicates the AGP obtain the notably superior performance than IPM as ${N_T}$ increases. In addition, the simulation is performed on an Intel Core i5-6200U CPU @ 2.3GHz 12GB RAM computer. As shown in Figure 4 (b), the average CPU time of the AGP is remarkably shorter than the IPM for solving the optimization problem of ($11$).

\section{Conclusion}
A low-complexity gradient projection approach has been proposed for solving the MIMO radar waveform design problem with CMC and SC constraints. By relaxing the feasible region, the key projection procedure is elaborately devised on the basis of the FISTA method. Numerical results reveal that the proposed AGP algorithm possesses a superior performance in terms of the SINR and the beampattern compared with the conventional IPM, with a much lower complexity.

\section*{Acknowledgement}
Authors wish to thank the support by the National Natural Science Foundation of China under Project No. 61771047.

\bibliography{mybibfile}

\end{document}